# From "Arbitrary Timberland" To "Skyline Charts": Is Visualization At Risk From The Pollution of Scientific Literature?

Lonni Besançon*
Linköping University

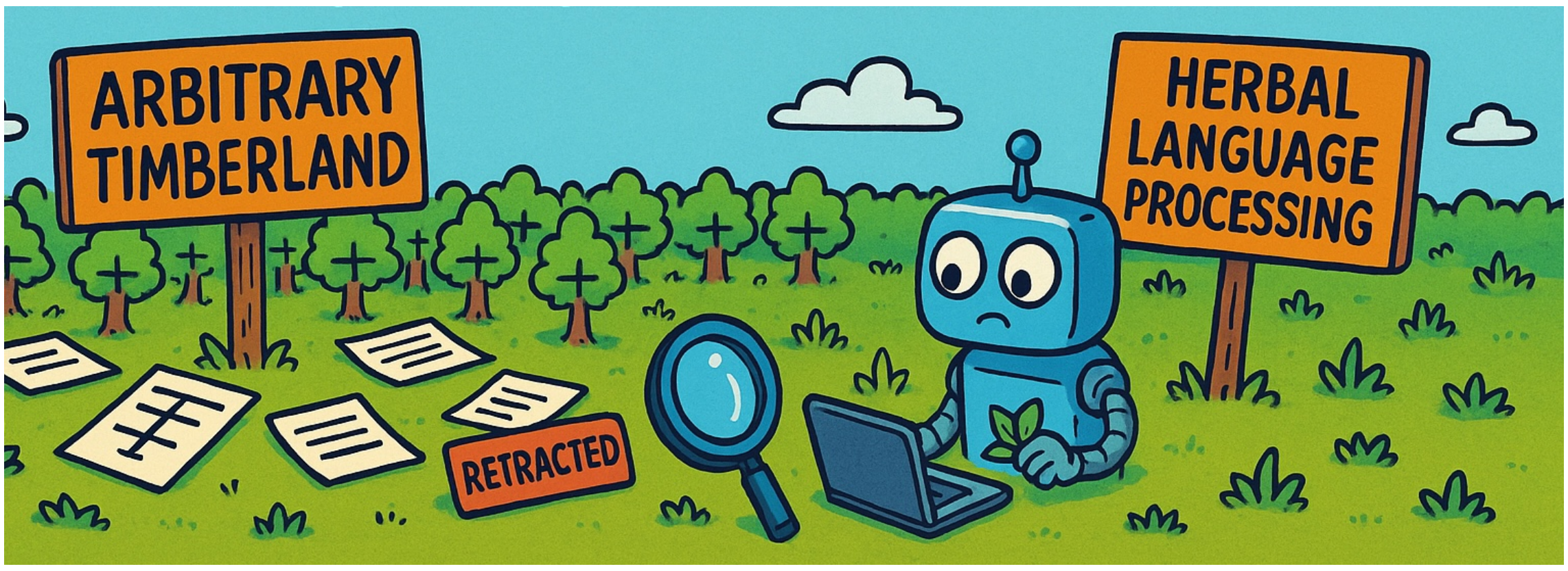

Figure 1: Abstract illustration of tortured phrases for visualization-related keywords. "Arbitrary Timberland" is a famous tortured phrase for "random forest", and "herbal language processing" for "natural language processing." Image generated by ChatGPT with human-in-the-loop cropping.

## Abstract

In this essay, I argue that, while visualization research does not seem to be directly at risk of being corrupted by the current massive wave of polluted research, certain visualization concepts are being used in fraudulent fashions and fields close to ours are being targeted. Worse, the society publishing our work is overwhelmed by thousands of questionable papers that are being, unfortunately, published. As a community, and if we want our research to remain as good as it currently is, I argue that we should all get involved with our variety of skills to help identify and correct the current scientific record. I thus aim to present a few questionable practices that are worth knowing about when reviewing for fields using visualization research, and hopefully will never be useful when reviewing for our main venues. I also argue that our skill set could become particularly relevant in the future and invite scholars of the fields to try to get involved.

## 1 Introduction

*Within this perspective manuscript, I contend that, although visual analytics inquiry does not appear to be immediately endangered by the present enormous surge of tainted scholarship, specific depiction notions are being exploited in deceitful manners and domains adjacent to ours are under assault. More troubling, the association disseminating our studies is inundated with myriad dubious manuscripts that are, regrettably, disseminated. As a collective, and provided we desire our scholarship to persist at the caliber it presently holds, I maintain that we ought to collectively engage our assortment of competencies to assist in spotting and amending the prevailing academic archive. I therefore intend to showcase several dubious behaviors that merit awareness when assessing domains incorporating visual analytics inquiry, and ideally shall never prove applicable when evaluating for our principal outlets. I likewise assert that our expertise may grow especially significant moving forward and encourage researchers in these disciplines to consider participating.*

If the previous paragraph appears difficult to parse for you: this is all absolutely normal. It actually is the abstract of this manuscript rewritten with a synonym of every single word by an LLM. This practice is unfortunately famous and has started to invade the scholarly record. It is widely known under the term "*tortured phrases*" [15]. They are the results of paraphrasing tools used to escape plagiarism detection" and are, unfortunately, only one of the many questionable practices we currently see in science and publishing processes. In the remainder of this paper, I want to raise awareness about some of these questionable practices and highlight the current threat it poses to science as a whole and perhaps to our field with a quick analysis of the current proportion of the scientific record being affected. I finish this essay with a touch of hope, mentioning how our skills can be helpful to detect and highlight problematic papers, also calling on all of my visualization colleagues to chime in and assist in the difficult task to try and maintain scientific integrity. The essay mostly focuses on my perception of these issues and does not aim to be exhaustive on the different forms of misconduct. The reason for writing an essay is, in the first place, to propose my experience and view as a structure to underline the problems that I mention here. Additionally, I am a strong believer of the fact that personal stories are more likely to be compelling, so here goes.

## 2 The Condition of the Craft of Dubious Behaviors
## The State of the Art of Questionable Practices

Questionable research practices are everywhere to be found. From citation gaming, to fake papers and including many other forms that

*e-mail: lonni.besancon@gmail.com

I will try to quickly summarize below. Most of them are linked to the proliferation of Paper Mills, companies that produce fake papers to boost one's researcher publication/citation index. Paper mills and their activities are a well-known and very documented problems of academia [1] but not all problematic research practices necesseraly revolve around them.

### 2.1 Mangled Wordings
### Tortured Phrases

Tortured phrases refer to the "results of paraphrasing tools used to escape plagiarism detection" [15]. They are mostly present in articles written by paper mills, companies paid to produce fake papers that simply copy-paste from existing papers and use a paraphrasing tool to avoid being flagged for plagiarism at the time of submission. The use of paraphrasing tool on known scientific expressions leads to nonsensical and frankly hilarious scientific terms. Examples include "invulnerable framework" for "immune system", "flag to clamor" for "signal to noise", "computerized reasoning" for "artificial intelligence", or "Lyme sickness" for "Lyme disease". One that I found recently is also particularly hilarious: "herbal Language Processing" for "Natural Language Processing". It is important to note that these tortured phrases and scientific expressions do not come from not being a native english speaker. While non natives can certainly write unfortunate sentences now and again, they tend to know what scientific expressions the field of research they are in uses. To detect these tortured phrases automatically Cabanac et al. [16] have developed the Problematic Paper Screener[1] that uses as fingerprints these tortured scientific expressions to scan the scientific literature and find papers that contain a few of these expressions before a human verifies the validity of the flag. As a matter of fact, because I mention many tortured phrases in this manuscript, it will likely be picked up by the Problematic Paper Screener, but the human validation aspect that goes with detecting papers containing those will make sure that the paper is not flagged as problematic, or so I hope.

### 2.2 Bibliographic Tactics
### Citation Gaming

Citation gamin is unfortunately common in academia and comes in different forms. From the rather widely known "cite my papers" in reviews to more obscure yet more damaging measures. The problem of reviewers asking only to cite their own paper is of course particularly important and supposedly quite prevalent considering the number of jokes around it in the scientific community. Yet, there are far more problematic practices when it comes to citation gaming. For instance, there are citation cartels, i.e., groups of individuals who excessively cite each other to artificially inflate their citation counts, even in cases when the papers are not relevant or even from another field [33]. Specific papers appear to be so-called "citation magnets", i.e., papers that are widely used to artificially inflate citation counts and tend to be found as references in a wide variety of papers outside of the field from which they should be referenced. As an example, I found a few papers in computer sciences almost exclusively referenced in chemistry papers. Other more obscure practices also exist. Whitefonting extra references consists in adding references to the PDF that are not cited in the main text of the paper in a white font so that they are not visible to the naked eye and only parsed by machines to increase citation counts [6]. Recently we have found evidence of "sneaked references" [10, 11], i.e., references present only in the metadata of paper submitted to crossref and not in the PDF of the article. This kind of fraud is new: so far fraudulent practices had been found in manuscript or reviews, but this was the first practice detected of metadata manipulation. In essence, the system relies on people working on the journal side to inject these additional references with or without the authors' knowledge. The tricky part in detecting sneaked references is that they are not found in the PDF of the article but only present in metadata submitted to crossref such that they would impact the citation dashboard that use crossref as a data source. One needs to manually compare the crossref database to the content of a PDF manually to detect those.

### 2.3 Synthetic Cognition Composition
### AI-Generated Content

The recent developments of LLMs also naturally impacted questionable research practices. A few papers have been found to contain direct phrases issued from a language model without disclosing the use of an LLM in the article. For instance, sentences such as "regenerate response" (resulting from an unfortunate and frankly clumsy copy-paste of the output of ChatGPT), "as of my last knowledge update in", or "as an language model I" have been found in published papers.[2] This practice does not stop at the generation of papers but is also used to generate reviews. Recently, authors have even been found to try and exploit this "lazy reviewer" vulnerability by injecting in their papers prompts for LLMs to only provide positive reviews of the paper.[3]

Of course, more subtle use of LLMs is quite difficult to detect although some approaches manage to find it in bulks of papers from a single journal by looking at the occurrence of specific words before and after the advent of ChatGPT [22]. One of the core issues here, is that non-native speakers could directly benefit, when writing scientific papers, from using LLMs to improve their writing [23, 24, 21] , such that it is difficult to distinguish when the LLM was used to help or when it was used to write the full article.

### 2.4 Misdirection of Ethical Clearance
### Ethical Approval

Ethics approval is a complicated topic that falls under both national and international laws and conventions. However, recent discoveries have pointed to inappropriate reuse of ethics approval for studies on human beings that are widely different [18]. Similarly, researchers have questioned whether some experiments on animal followed the proper ethical guidelines.[4] These topics are difficult to address as laws are rarely written in English and ethics application processes are often opaque to outside researchers.

### 2.5 Pictorial redundancy
### Image Duplication

Another core issue is the reuse of images or parts of images in papers. In some cases, the images are reused within different groups in the same article, in others, they are used in two totally different papers to show different things. In most cases, this is done without acknowledging the existence of the initial figure, and more worryingly, images are also slightly modified (rotated, altered, cropped, ...) to avoid detection. This practice is largely documented [13] and AI tools such as *ImageTwin.ai* can assist in finding those.

## 3 Is the International Visual Data Discourse Symposium Confronting a Potential State of Existential Peril?
## Is IEEE VIS at Risk?

So far, and to the best of my knowledge, none of these practices have been found in IEEE VIS papers. This is probably due to the fact that our community is particularly small and if such fraudulent practice was found from one of us the impact would be devastating.

[1] https://dbrech.irit.fr/pls/apex/f?p=9999:1::::::

[2] https://retractionwatch.com/papers-and-peer-reviews-with-evidence-

[3] https://www.theguardian.com/technology/2025/jul/14/scientists-reportedly-hiding-ai-text-prompts-in-academic-papers-to-re

[4] https://scienceintegritydigest.com/2020/05/07/animal-ethics-misconduct-mice-with-very-large-tumors/

This is not so much the case for communities of a few dozen thousands of researchers where one questionable practice can easily be forgotten or not caught. I also would argue that our peer reviewing would tend to be quite thorough in HCI and data visualization and that such practices would immediately be caught. This is reassuring of course but it should not prevent us from being vigilant. In particular I would like to point to specific points that I believe we, as a community, should be particularly paying attention to.

About Tortured Phrases, although they do not seem to be present in IEEE VIS papers or IEEE TVCG papers at all, they seem to permeate in IEEE publication. A recent scan with the problematic paper screener revealed 47581 papers published by IEEE containing problems found by the Problematic Paper Screener, [5], of those, 6583 are papers containing tortured phrases. I have flagged of few hundred of those on Pubpeer but so far most of them remain publisher despite the obvious use of a paraphrasing tool. Funnily enough some of the tortured phrases pertain to the field of visualization. One example that particularly strikes me is the use of "arbitrary backwoods" or "irregular timberland" for "random forest" plots. While there is little doubt that our community wouldn't catch such expression, we should perhaps make sure to report more thoroughly the papers that our main publisher does not seem to catch.

I am not particularly worried about questionable practices on citations either to be honest. Again, the size of our field alone is an argument in and on itself to prevent anyone from trying such practices. Similarly, I am not particularly worried about the ill-intended use of LLM in our field. However, I believe that we should already anticipate the impact and role that LLM can have on our writing. It is known that native language speakers may have an advantage when it comes to getting their papers accepted [30, 34], and that in general being a native english speaker is an advantage in academia [2]. It is therefore not unreasonable to think that one author may want to use an LLM to polish their writing and thus increase the fluidity of their manuscript and the quality of the paper [23]. In fact, before the recent advances of LLMs, many in the field relied on Grammarly to improve their writing and did not have to disclose its use in the manuscript. We should therefore be thinking ahead of the use cases that are acceptable or not when it comes to LLM and make sure that our policy does not become unfair for non native speakers.

Beyond those cases, I believe our field can do better when it comes to ethics approval and their mention in papers. Again, national and international laws are complicated and difficult to verify for non-native speakers. However, we should be making an extra effort towards transparency in this matter, following current practices in other field: mention ethics approval number and where it was obtained from, share why the study was exempt from ethics approval, etc. Although some papers in our field already do so, this is still not a common practice of the community. As a matter of fact, after I flagged with my co-authors ethics approval concerns in biomedical research from a single institute in a paper [18] and through Pubpeer or the media, there has seem to be some retaliations on my own VIS and HCI papers on this matter. I would thus argue that the IEEE VIS community strongly advocates for more transparency on this topic in all of our published research papers, giving space in our template for the mention of such matters.

## 4 So what's the point of this paper?

I decided, earlier this year, to write this paper for two main reasons. First, I want to raise awareness about these problems that are, unfortunately for science and its trustworthiness, quite pervasive and difficult to combat. Doing so, on my side, came at a great cost and some risks for my family and career [31], but the analyses that I have provided with my co-authors have helped shaped better methods across different fields of research, including even a revision of the Declaration of Helsinki [28] for biomedical research ethics, or a push for more transparency of research papers after we had found that hundreds of COVID-19 papers were peer-reviewed in a single day or less with, in around 30% of cases, editorial conflicts of interests [12]. As I have advocated many times, this analysis work is absolutely possible to do while doing our regular visualization research and I hope to get more of the community to join in on this effort to clean up the scientific records. Science is, overall, still completely trustworthy and robust, but the pollution/contamination of our literature [14] is a danger to our credibility first, but also to our modern societies, as exemplified recently by the recent *skepticism* (ahem...) towards scientific advances and institutions from the new US administration. I also wish to raise awareness about the importance of this work which *should* be considered a part of our job description. Too often, peer-review is unfortunately restricted to pre-publication peer-review when it should actually also include post-publication peer review done on platforms such as Pubpeer [3], or directly through the media or social media [20].

Second, I would like to highlight that our skillset, as computer-science majors and visualization and data analysis researchers is also crucial to this work. Recently, I have proposed, along with my co-author Fabrice Frank to create a visual dashboard for editors and publishers to see which papers, flagged as problematic, are getting traction and attention such that they could focus their research integrity efforts towards those first [17]. This is particularly important if we consider that some papers that are problematic or even retracted have, in the past, shaped policies, in particular when it comes to public health. But beyond this example, we have a skillset for data analysis, experimental design, controlled studies, or even statistics, that can be helpful in analyzing studies in other fields too. In 2021, along with Raphael Wimmer, another HCI researcher, we managed to get a flawed paper on COVID-19 retracted [32] by simply using the code made available and playing with it to show that the authors were actually wrong in their methods [25]. Similarly, by working with epidemiologists, statisticians, and medical doctors, we managed to publish our methodological, ethical, and reviewing concerns [4] about the study that started the hype on hydroxychloroquine as a cure for COVID-19, retracted 4 years later. While these are only examples, they clearly highlight that our skills are very useful outside of our community and that we can do this work by simpling reaching out to people who are interested in correcting the academic record.

Third, I think it is important to raise awareness that "sleuths" (as people trying to correct the scientific records are often called) come from a diversity of background but are often on the more junior side of the academic career path [**?**]. As such, I have faced numerous difficulties and dangers (for me, my career, or my family [31]) that echo the experience of other sleuths [29] (or COVID-19 researchers for that matter [27]). It is important to raise awareness on this topic and for other scientists to understand what usually comes with sleuthing work: tentative retributions, slandering emails, smearing campaigns [29, 31]... and to try and force institution to protect academic post-publication peer review [7].

Finally, it may be time for us to consider the future of scientific publishing beyond the traditional submit-review-publish we currently know. This model is particularly problematic because papers found to contain problems rarely get timely corrections or retractions [5, 9]. In my many efforts to get corrections or retractions, even on papers that pertained to public health and directly impacted policy making during COVID, it always took several months to get any form of editorial action on problems [25, 26]. The inertia of publishers to change their practices is particularly problematic. Too few journals have tried to change this paradigm, with eLife the only major journal pushing strongly in this direction. In our field, Chat

[5] https://dbrech.irit.fr/pls/apex/f?p=9999:3::::RIR:IRC_PUBLISHER:IEEE

Wacharamanotham, Florian Echtler, Matthew Kay, and myself are trying to change things and experiment with different submission, reviewing, and publishing models through the Journal of Visualization and Interaction JoVI [8]. Speaking for myself, I do not think our goal is to replace IEEE VIS or IEEE TVCG but rather to propose something different. The scientific method, processes and publications have been in constant evolution for centuries, we should continue to question ourselves, our methods, and our practices and make sure this introspection is thoroughly done. It would show our honesty despite what some government today will try to say of it. As a matter of fact, I am happy to report that thanks to a lot of verification efforts from both critics, friends, and journalists, one of my own paper was found to have some very minor problems in its data, which have now earned the initial paper a correction [19]. This was handled very professionally and quickly by the editor in chief of the journal we published in and I am grateful for this. Not only did we manage to correct some errors (that did not in any way impact our findings), but we also showed that, despite the label that we too often get attributed of "sleuths trying to correct science," we apply this introspection to ourselves too.

## 5 Conclusion

I figure it is now time for me to answer the question in my own title here. "Is visualization at risk from the pollution of the scientific literature?" As one may expect, following the *Betteridge's law of headlines*,[6] the answer I would give is "No." But there is a lot we can do as (visualization) researchers to ensure it remains the case and to help our colleagues across all fields of science correct the scientific record and preserve it.

## Acknowledgments

I wish to thank my sleuthing friends and colleagues who have supported me in this work as well as Digital Science who is giving me support to build the tool VIRUS with Fabrice Frank. For references of papers published by Springer Nature which does not offer an export to bibtex function, the references were formatted by ChatGPT.

[6] https://en.wikipedia.org/wiki/Betteridge%27s_law_of_headlines